# Topological susceptibility and Instanton size distribution from over-improved cooling

Philippe de Forcrand[a] and Seyong Kim[b]

[a]IPS, ETH-Zürich, CH-8092 Zürich, Switzerland

[b]Center for Theoretical Physics, Seoul National University, Seoul, Korea

We measure the topological susceptibility by cooling with an over-improved action. In contrast with usual cooling, large instantons survive over-improved cooling *indefinitely*. By varying the parameter of the over-improved cooling action, we measure the instanton size distribution.

## 1. Motivation

A variety of methods are available to measure the topological charge $Q$ of a non-abelian lattice gauge field. To choose among them, one is guided by generally conflicting goals:
1) the measurement of $Q$ should be insensitive to "dislocations", short-range fluctuations caused by the lattice discretization;
2) $Q$ represents a topological index, and thus should be an integer;
3) the measurement of $Q$ must be "objective": it must not depend on a tuning parameter whose value is fixed by subjective criteria.

The most common methods, "geometric", "analytic" and "cooling", miss at least goals 1, 2, and 3 respectively. Here we try to address the deficiencies of the latter.

Cooling (or smearing) consists of an iterative, local minimization of the Wilson action. Cooling dynamics can be slowed down if desired, by under-relaxation [1]. Cooling has become widely used to gain qualitative insight into the large-scale features of the gauge field. Under cooling dynamics, many observables show metastability: their value reaches a plateau before the configuration eventually collapses to the trivial vacuum. How soon this plateau is reached and how long it lasts depends, however, on the observable, and even more problematically, on the initial gauge configuration. This lack of robustness has prevented widespread acceptance of cooling as a quantitative method.

The behavior of instantons under cooling is well-known. As the lower curve of Fig.1 shows, the action of a lattice instanton decreases monotonically with its size. Thus in the process of minimizing the action, cooling will shrink instantons until their size becomes $\mathcal{O}(a)$: then the lattice discretization is so coarse that the non-trivial topology of the gauge-field cannot be detected, even by a geometric method. The instanton has "dropped through the lattice". When one looks for a plateau in the topological charge $Q$ under cooling, one tries to observe the slow shrinking of the largest instantons. It is not at all clear how many smaller, physical instantons have already disappeared at this stage.

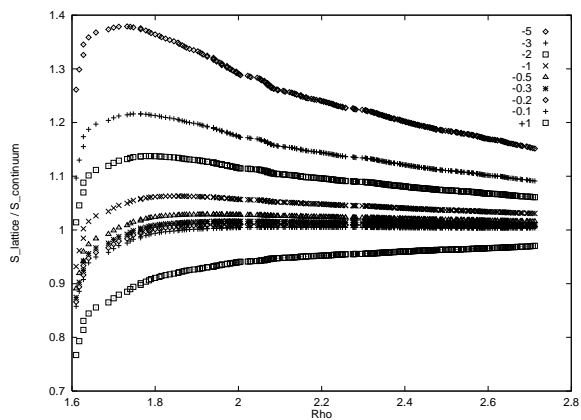

Figure 1. Action of a lattice instanton versus its size, for the Wilson action (bottom curve) and various over-improved actions $S(\epsilon)$.

To remedy this situation we propose to use over-improved (OI) cooling [2]. An over-improved action, at tree-level, makes positive the first term $\mathcal{O}((a/\rho)^2)$ in the instanton action: so the action of large instantons now increases as they shrink. Since by construction the lattice instanton action must go to zero with its size, the action will develop a maximum for a size $\rho_{peak}$. The upper curves of Fig.1 show this effect for various values of the over-improvement parameter $\epsilon$. Minimizing this action by cooling will then shrink instantons of size smaller than $\rho_{peak}$ until they disappear; those larger than $\rho_{peak}$ will expand to the maximum size allowed by the finite lattice. They will be stable against any number of cooling steps. Thus we can consider an *objective* procedure to measure the net topological charge contributed by instantons of size $\rho > \rho_{peak}$: cool with an over-improved action until a stable state is reached; measure $Q$ then. This measurement is facilitated by the fact that instantons are now maximally large and smooth: the simplest, analytic method (measuring $\int F_{\mu\nu}\tilde{F}_{\mu\nu}$) can be employed. By varying the over-improvement parameter $\epsilon$, and thus $\rho_{peak}(\epsilon)$, the size distribution of instantons can be investigated.

## 2. Results

For economic reasons, we test our approach on $SU(2)$ rather than $SU(3)$. We use the action [2]:
$$S(\epsilon) = \frac{4-\epsilon}{3}Tr(\mathbf{1}-W_{1\times1}) + \frac{\epsilon-1}{48}Tr(\mathbf{1}-W_{2\times2}) \quad (1)$$
where $\epsilon < 0$ for over-improvement. This action has been used so far to produce instantons in situations where they exist in the continuum, ie. with twisted boundary conditions. Here we want to use it with periodic b.c., where no continuum instanton exists. We were therefore concerned that our stable state after OI cooling might be distorted from a continuum instanton. Fig.2 shows the distribution of the topological charge density away from the instanton center, measured by $F_{\mu\nu}\tilde{F}_{\mu\nu}$, where $F_{\mu\nu}$ is extracted from the OI action (1), on a $16^4$ lattice after (infinitely) long cooling with $\epsilon = -1$. The solid line is t'Hooft's continuum ansatz $\propto (\frac{\rho^2}{x^2+\rho^2})^4$ [periodic images have a negligible effect on the ansatz]. The fit is excellent up to large distances where the boundary of the box is felt. The total charge and total action are very close to their continuum values (the maximum deviation was $\sim 7\%$ on the smallest $8^4$ lattice). The surprise however is that the instanton size does not exceed $\rho/a \simeq 3.5$, even on a $16^4$ lattice. We consistently observed a maximal size $\rho_{max}/a < L/4$ in all our simulations. This can be contrasted with much larger sizes observed with twisted b.c. (see eg. [3]). Thus periodic b.c. induce very strong finite size effects on instantons.

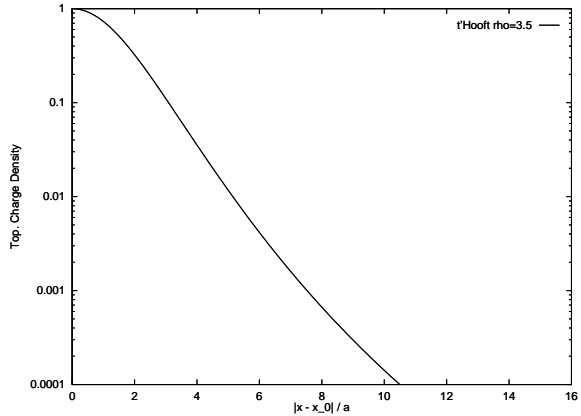

Figure 2. Profile of the topological charge density for the maximum size instanton ($L = 16, \epsilon = -1$). The solid line is t'Hooft's ansatz with $\rho/a = 3.5$.

We kept periodic b.c. nonetheless, to compare our results with previous work [1]. Our procedure consisted then of generating a sample of configurations with the Wilson action, and of cooling each of them several times, with OI actions of successively more negative $\epsilon$. If $n(\rho)$ is the number of instantons of size $\rho$, we measure on each configuration, after (infinitely) long OI cooling:
$$Q(\epsilon) = \int_{\rho_{min}(\epsilon)}^{\rho_{max}(L)} d\rho \; n(\rho) \quad (2)$$
The upper bound $\sim L/4$ in the integral comes from the finite volume as just discussed. The lower bound is the threshold size for instanton survival under OI cooling: it is $\rho_{peak}(\epsilon)$ as visible on Fig.1. Then from the distribution of $Q(\epsilon)$ over our sample, we extracted an effective susceptibility
$$\chi_L(\epsilon) = \; <Q^2(\epsilon)> \; / \; V \quad (3)$$

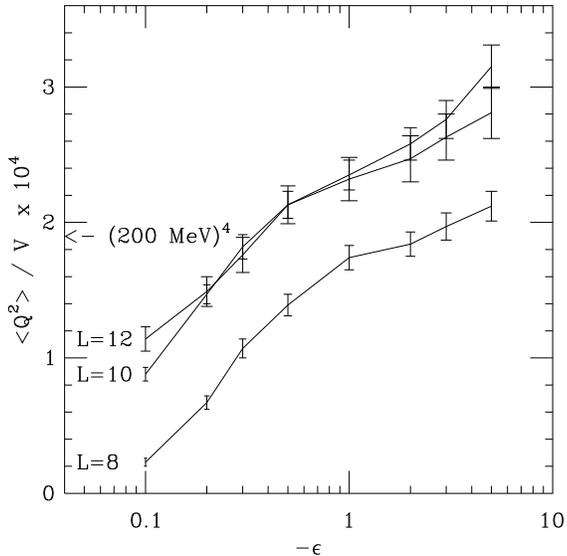

Figure 3. Topological susceptibility as a function of $\epsilon$, for various lattice sizes ($\beta = 2.4$).

Our method is expensive: we wanted to make sure that we reached a stable state, so we cooled with 100-700 sweeps for each value of $\epsilon$. This is the price we pay for an objective method.

Our sample consisted of 1000 (540) configurations for $L = 8, 10, (12)$ at $\beta = 2.4$, separated by 50 Monte Carlo sweeps. Remarkably, no measurable autocorrelation could be detected in $Q(\epsilon)$, even for $\epsilon = -0.1$: the largest instantons appeared as drawn from a random sequence. Our MC update was a 9-to-1 mixture of over-relaxation and heatbath. Preliminary tests with a pure heatbath update revealed no autocorrelation either. This is a welcome finding on the efficiency of pure gauge MC updates, in sharp contrast with the long autocorrelation times observed for $Q$ with dynamical fermions, using Hybrid Monte Carlo [4].

Our results for the effective susceptibility (3) are shown in Fig.3. Several observations are in order:
1) as $\epsilon$ is changed, the lower cutoff $\rho_{min}(\epsilon)$ is changed. The effect on the susceptibility is dramatic (a factor $\sim 10$ on the $8^4$ lattice). If we could push $\rho_{min}$ to zero, presumably we would recover the susceptibility obtained via the geometric method. It is clear that measurements of the topological susceptibility are meaningless if not accompanied by an estimate of the lower cutoff $\rho_{min}$.

Ideally $\rho_{min}$ should be adjusted to include all physical instantons but leave out all lattice artifacts. Although an estimate $\rho/a \sim 1.93$ for this ideal value can be obtained [3], it really should be extracted a posteriori, by looking at the scaling with $\beta$ of the instanton size distribution: lattice artifacts will give a divergent contribution below $\rho_{min}$, while the distribution of real instantons above $\rho_{min}$ will scale.
2) as $L$ is changed, the upper cutoff $\rho_{max}(L)$ is changed, more or less in proportion to $L$. Therefore the difference $\chi_{L_1}(\epsilon) - \chi_{L_2}(\epsilon)$ should be independent of $\epsilon$: it comes from large instantons, which fit inside the large box but not the small one. Fig.3 shows clearly that this is the case: the data for $L = 8$ lie below the rest, with a nearly constant offset. On the other hand there is little difference between $L = 10$ and $12$, indicating that the contribution of large instantons of size $> \rho_{max}(10) \sim 2.3$ to the true, infinite volume susceptibility is very small.
3) using $a^{-1}(\beta = 2.4) = 1.7 GeV$, an arrow marks the phenomenologically plausible susceptibility $(200 MeV)^4$. Such a result could be obtained with $\epsilon \sim -0.3$, which is the value favored in [5].

The next step consists of mapping our results as a function of $\rho_{min}$, not $\epsilon$. $\rho_{min}(\epsilon)$ is the average minimum size of an instanton, generated by the Wilson action, necessary for survival under $OI(\epsilon)$ cooling. Here we measured $\rho_{min}(\epsilon)$ by monitoring $S(\epsilon)$ evaluated on a shrinking instanton, being cooled with the Wilson action. The size was determined in a manner similar to [1], by looking for a hypersphere containing half the total charge. One difficulty is that an instanton for the Wilson action is not an instanton for the OI action (it does not satisfy the classical equations of motion): so different Wilson instantons of the same size may suffer different fates under $OI(\epsilon)$ cooling. In addition, $\rho_{min}$ also depends on the proximity of the instanton to a lattice site. An attempt at averaging is performed in [3]. Here we merely acknowledge these effects as a source of uncertainty.

Extracting $\rho_{min}(\epsilon)$ from Fig.1, the effective susceptibility is displayed in Fig.4 as a function of $\rho_{min}$. The value $(200 MeV)^4$ corresponds to $\rho_{min}/a \sim 1.96$, very close to the estimated cutoff for lattice artifacts $\sim 1.93$ [3]. We also present cubic fits (higher degree polynomials are not constrained enough): the maximal instanton sizes given by these fits ($\sim 2.25$ and $\sim 2.35$) are consistent with the sizes actually measured after OI cooling.

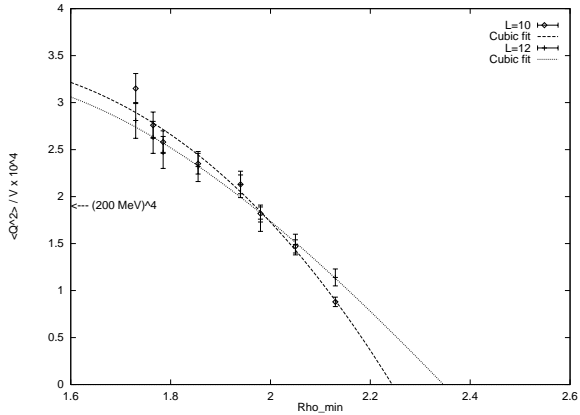

Figure 4. Effective susceptibility versus minimum instanton size.

Finally we can extract the instanton size distribution. Since $Q^2(\epsilon) = \int d\rho_1 d\rho_2 n(\rho_1) n(\rho_2)$, and assuming that instantons interact only weakly, so that correlations between $n(\rho_1)$ and $n(\rho_2)$ are negligible, we get
$< Q^2(\epsilon) > = \int_{\rho_{min}(\epsilon)} d\rho < n^2(\rho) >$,
so that
$< n^2(\rho_{min}) > = \frac{d}{d\rho_{min}} < Q^2 >$.
The differentiation of our cubic fit ($L = 12$) is shown in Fig.5. It is striking that, although allowed by the degree of the polynomial fit, and unlike all previous studies, no maximum appears in the instanton size distribution. Several explanations are possible:

• we have large statistical uncertainties, especially in the determination of $\rho_{min}(\epsilon)$.
• our assumption $< n(\rho_1) n(\rho_2) > = \delta(\rho_1, \rho_2)$ may be flawed because of (anti-)instanton interactions, non-negligible in such a small box.
• all other studies measure instanton sizes on cooled configurations. At that stage, small instantons have already shrunk and disappeared, and small, close-by A-I pairs have annihilated. Measuring the size distribution at that stage causes a bias, which depletes the small size population and may be responsible for the observed peak. Our procedure avoids this bias.

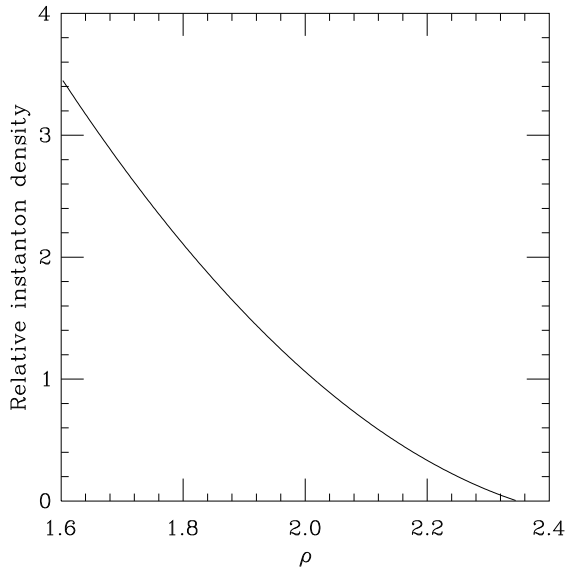

Figure 5. Fitted instanton size distribution $< n^2(\rho) >$ ($L = 12$).

We thank I.O. Stamatescu and especially Margarita García Pérez who provided the data of Fig.1.